\documentclass[letterpaper, 10 pt, conference]{ieeeconf}

\IEEEoverridecommandlockouts
\usepackage{cite}
\usepackage{graphics} 
\usepackage{epsfig} 
\usepackage{mathptmx} 
\usepackage{times} 
\usepackage{amssymb}  
\usepackage[cmex10]{amsmath}
\usepackage{amsthm}
\usepackage{amsfonts}
\usepackage{latexsym}
\usepackage{mathrsfs}
\usepackage{upgreek}
\usepackage{xcolor}
\usepackage[normalem]{ulem} 
\usepackage{mathtools}
\usepackage{xspace}
\usepackage{geometry}
\usepackage{graphicx}
\usepackage{enumerate}
\usepackage{subcaption}

\usepackage{algorithm}
\usepackage{algpseudocode}

\newcommand{\rulesep}{\unskip\ \vrule\ }

\usepackage[linkbordercolor={1 1 1},citebordercolor={1 1
  1},urlbordercolor={0.0 0.0
  0.0},urlcolor=blue,colorlinks=true,linkcolor=black,citecolor=black]{hyperref}
\usepackage{url}
\usepackage{cleveref}

\theoremstyle{remark}
\newtheorem{rem}{Remark}
\theoremstyle{definition}
\newtheorem{defn}{Definition}

\newtheorem{example}{Example}
\theoremstyle{plain}

\newtheorem{thm}{Theorem}

\newcommand{\R}{\mathbb{R}}
\newcommand{\A}{\mathcal{A}}
\newcommand{\udes}{u_{\textrm{des}}}
\newcommand{\uact}{u_{\textrm{act}}}
\renewcommand{\S}{\mathit{S}}

\title{\LARGE \bf
Safe Drone Flight with Time-Varying Backup Controllers
}

\author{Andrew Singletary, Aiden Swann, Ivan Dario Jimenez Rodriguez, and Aaron D. Ames
\thanks{
Andrew Singletary, Aiden Swann, and Aaron D. Ames are with Department of Mechanical and Civil Engineering,
        Ivan Dario Jimenez Rodriguez with Deparment of Computational Math and Science,
        California Institute of Technology, Pasadena CA 91125, U.S.A. Email addresses:
		{\tt \small \{asinglet, aswann, ivan.jimenez, ames\}@caltech.edu}. This work is supported by AeroVironment,  NSF CPS award \#1932091, and BP.
}
}

\begin{document}

\maketitle
\thispagestyle{empty}
\pagestyle{empty}

\begin{abstract}
The weight, space, and power limitations of small aerial vehicles often prevent the application of modern control techniques without significant model simplifications.
Moreover, high-speed agile behavior, such as that exhibited in drone racing, make these simplified models too unreliable for safety-critical control.
In this work, we introduce the concept of time-varying backup controllers (TBCs): user-specified maneuvers combined with backup controllers that generate reference trajectories which guarantee the safety of nonlinear systems.
TBCs reduce conservatism when compared to traditional backup controllers and can be directly applied to multi-agent coordination to guarantee safety.
Theoretically, we provide conditions under which TBCs strictly reduce conservatism, describe how to switch between several TBC's and show how to embed TBCs in a multi-agent setting.
Experimentally, we verify that TBCs safely increase operational freedom when filtering a pilot's actions and demonstrate robustness and computational efficiency when applied to decentralized safety filtering of two quadrotors.
\end{abstract}

\section{INTRODUCTION}
\label{sec:introduction}



Drones have emerged as a staple robotic technology that have made the transition to the general public, used by hobbyists and videographers, and are finding application in a variety of domains.  
Contributions from the robotics research community (see \cite{mellinger2011minimum,mueller2014stability,sreenath2013geometric,hwangbo2017control} to name a few) led to this transition to practice, and drones remain an active area of research due to their availability, small size and weight, nonlinear dynamics with important controllability properties, e.g., differential flatness \cite{faessler2017differential}, and increases in mobile computation.  The result is the demonstration of impressive feats of agility, but these generally require either significant offline computation \cite{kaufmann2020deep}, or use computationally expensive predictive controllers \cite{loquercio2021learning,sun2021comparative}. 
Lacking from these successes in acrobatic flying are guarantees of safety.  

Safety concerns and regulations have thus far prevented the wide-scale adoption of applications that involve close proximity to humans, e.g., drone delivery.  Due to their significant kinetic and potential energy, even a small drone can pose a life-threatening risk when operated incorrectly.  This leads to the concept of regulating an operator's desired input to ensure safety which has been explored in several contexts.  In \cite{broad2019highly}, the authors leverage GPUs to generate thousands of potential trajectories via MPC, and utilize the closest to the desired input. In \cite{tearle2021predictive}, the authors present a promising MPC-based approach to safety filtering for vehicle racing, but the method has not shown to be applicable to higher-dimensional systems with less computational power.

\begin{figure}
    \centering
    \includegraphics[width=\columnwidth]{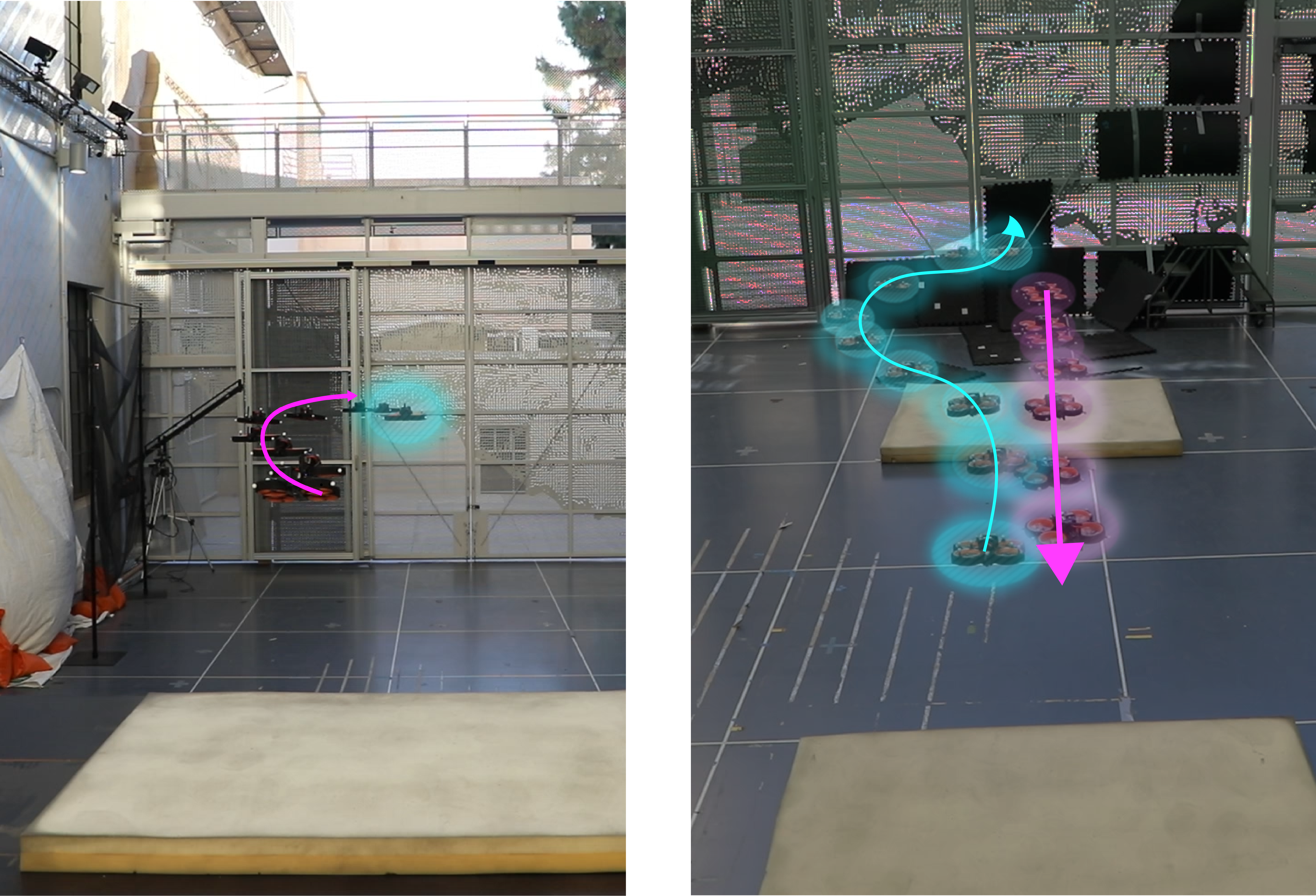}
    \caption{Illustration of the experimental results demonstrating safe flight of multiple drones through the use of time-varying backup controllers.}
    \label{fig:intro}
\end{figure}

Control barrier functions \cite{ames2017cbf} have been widely utilized in recent years due to their computational efficiency and their ability to handle general nonlinear dynamical systems of arbitrary dimension. They have been utilized in a wide variety of quadrotor applications \cite{wu2016safety,khan2020barrier,xu2018safe}, but none of these results were computed online on flight hardware. Multi-agent drone systems have also been considered in the context of control barrier functions, as shown in \cite{wang2018safe}, albeit utilizing simplified model dynamics and centralized computing. In \cite{chen2020guaranteed}, the authors present a decentralized approach to safety filtering of multiple agents utilizing the shared knowledge of backup control policies that keep the individual agents safe, but this was only tested in simulation, and could not be computed on flight hardware.


This paper presents a new approach for achieving flight with formal guarantees of safety that can be realized in practice onboard highly dynamic drones.  
To this end, we build upon previous work \cite{singletary2022onboard}, wherein the authors introduce a method for utilizing backup controllers to guarantee safety in a way that did not require solving an optimization problem or computing expensive gradients online. In this work, we extend the method to include multi-agent aerial systems, as well as present a new framework for drastically increasing the operational freedom of the user through the use of maneuvers.  The notion of maneuvers is formally encoded by the concept of \emph{time-varying backup controllers (TBCs)} which formally guarantee safety.  We establish that theoretically time-varying backup controllers produce larger safe sets than backup controllers without this time-varying component, thereby giving the operator more control authority without sacrificing safety.  These theoretic results are demonstrated in simulation and on hardware in the context of collision avoidance, both for a single agent and for two agents.  The computational benefits of the approach are illustrated through the online implementation, and the theory is validated through the resulting safe behaviors. 

The layout of the paper is as follows. 
Section \ref{sec:motivation} overviews the underlying mathematical principles we will use to provide safety guarantees.
Section \ref{sec:theory} showcases Time-varying backup controllers, the algorithms and methods used to practically implement them and the conditions under which they provide safety guarantees with no added conservatism. We also show how to apply TBCs to multi-agent systems in a decentralized manner.
Section \ref{sec:implementation} outlines how to generate and implement these backup maneuvers, and demonstrates several of the concepts on a quadrotor in simulation.
Finally, Section \ref{sec:results} demonstrates these ideas with a hardware implementation of two quadrotors with only small microprocessors for computation.

\section{Preliminaries}
\label{sec:motivation}
\subsection{Safe flight and set invariance}

We begin by reviewing safety as set invariance. 
Consider a general nonlinear control-affine dynamic system:
\begin{align} \label{eqn:dyn}
    \dot{x} = f(x) + g(x) u,
\end{align}
with state $x \in \mathbb{R}^{n}$, input $u \in U$ for the admissible input set $\mathit{U} \subseteq 
\mathbb{R}^{m}$.

For a policy $u = \pi(x)$ and initial condition ${x(t_0) = x_{0} \in \mathbb{R}^{n}}$, the solution to this closed-loop system is given by the flow map 
\begin{equation}
    {x(t)=\Phi_{\pi}(x_{0},t)} = x_0 + \int_{t_0}^t \left( f(x) + g(x) \pi(x) \right)dt , \ t \geq t_0.
\end{equation}

Consider $\S$ to be the set of all states in which the system is safe.
In this work, we consider the set of all states where the drones are not in collision.
We introduce the notion of set invariance with the goal of ensuring that our drone stays in this set for all time without being overly restrictive on the actions the drone takes inside this $\S$.

\begin{defn}[Set Invariance]
A set $\mathit{S}$ is called \textit{invariant} if the system's state stays in $S$  for all time, i.e. $\forall t \geq t_0,\ x(t) \in \mathit{S}$. 
\end{defn}

We restrict our attention to safe sets defined by the 0-level set of a continuously differentiable function as follows:

\begin{defn}(Safe Set)
Let $\S \subset \R^n$ be the set defined by a continuously differentiable function $h: \R^n \to \R$:
\begin{eqnarray}
\S & = & \{ x \in \R^n ~ : ~ h(x) \geq 0 \} , \nonumber\\
\partial \S & = & \{ x \in \R^n ~ : ~ h(x) = 0 \}, \nonumber\\
\mathrm{Int}(\S) & = & \{ x \in \R^n ~ : ~ h(x) > 0 \}. \nonumber
\end{eqnarray}
\end{defn}

\subsection{Constructing Safe Sets with Backup Controllers}

Suppose we have a subset of the state-space we wish to render forward invariant $S = \{ x \in \mathbb{R}^n | h(x) \geq 0 \}$.
Furthermore we assume there exists a set $S_B = \{ x \in \mathbb{R}^n | h_B(x) \geq 0 \}$ that is \textit{control invariant} as follows:

\begin{defn}[Control Invariant Set]
Safe set $S$ is \textit{control invariant} if there exists a control policy $\pi : \mathbb{R}^n \rightarrow U$ that renders $S$ invariant.
\end{defn}

We assume that this $\S_B \subset \S$ and it usually corresponds to a trivially controllable subset of the system's state-space.
We will call the particular control invariant set $\S_B$ the \textit{backup set}.
In the case of a drone, this could correspond a small region around a steady hover.
The set $\S_B$ can be significantly smaller than $\S$, and the size of $\S_B$ has almost no effect on the performance of this method.

As shown in \cite{chen2021backup}, given a control policy $\pi$ we can define a corresponding implicit control-invariant set:
\begin{align}
    &S_I(\pi) = \nonumber\\ 
    &\left\{ 
    x_0 \in \R^n \left| \underset{\forall t \in [0,T]}{\Phi_{\pi}(x_0,t)} \in \S \wedge \Phi_{\pi}(x_0,T) \in S_B\right.
    \right \} \label{eq:si_set} =\\
    &\left\{ 
    x_0 \in \R^n \left| \min_{t \in [0,T]} \left\{h(\Phi_{\pi}(x_0,t)),  h_B(\Phi_{\pi}(x_0,T))\right\} \geq 0\right.
    \right \} \label{eq:si_min}
\end{align}

In other words, this implicit safe set corresponds to the set of initial conditions such that flows of the system under the policy $\pi$ remain in $\S$ for $t \in [0,T]$ and reach $S_B$ by time $T$. We call policies with a non-empty $S_I$ \textit{backup controllers}.

This particular construction allows for the creation of an \textit{implicit control barrier function (ImCBF)}:

\begin{defn}[Implicit control barrier function (ImCBF)]
The flow map of a system $\Phi_{\pi}$ with associated backup controller $\pi$ defines an \textit{implicit control barrier function(ImCBF} $h_I$ defined as:
\begin{align}
    h_I(x) \coloneqq \min_{t \in [0,T]} \left\{h(\Phi_{\pi}(x_0,t)),  h_B(\Phi_{\pi}(x_0,T))\right\}
\end{align}
So that $h_I$ defines the set $S_I(\pi) = \{x \in \mathbb{R}^n \left| h_I(x) \geq 0 \right.\}$.
\end{defn}

While these ImCBF's can be used in the context of a CBF-based QP controller, as demonstrated first in \cite{gurriet2018online}, this implementation is not practical for small aerial vehicles with limited computational power due to the expensive gradient computations involved.

\begin{figure}
    \centering
    \includegraphics[width=.9\columnwidth]{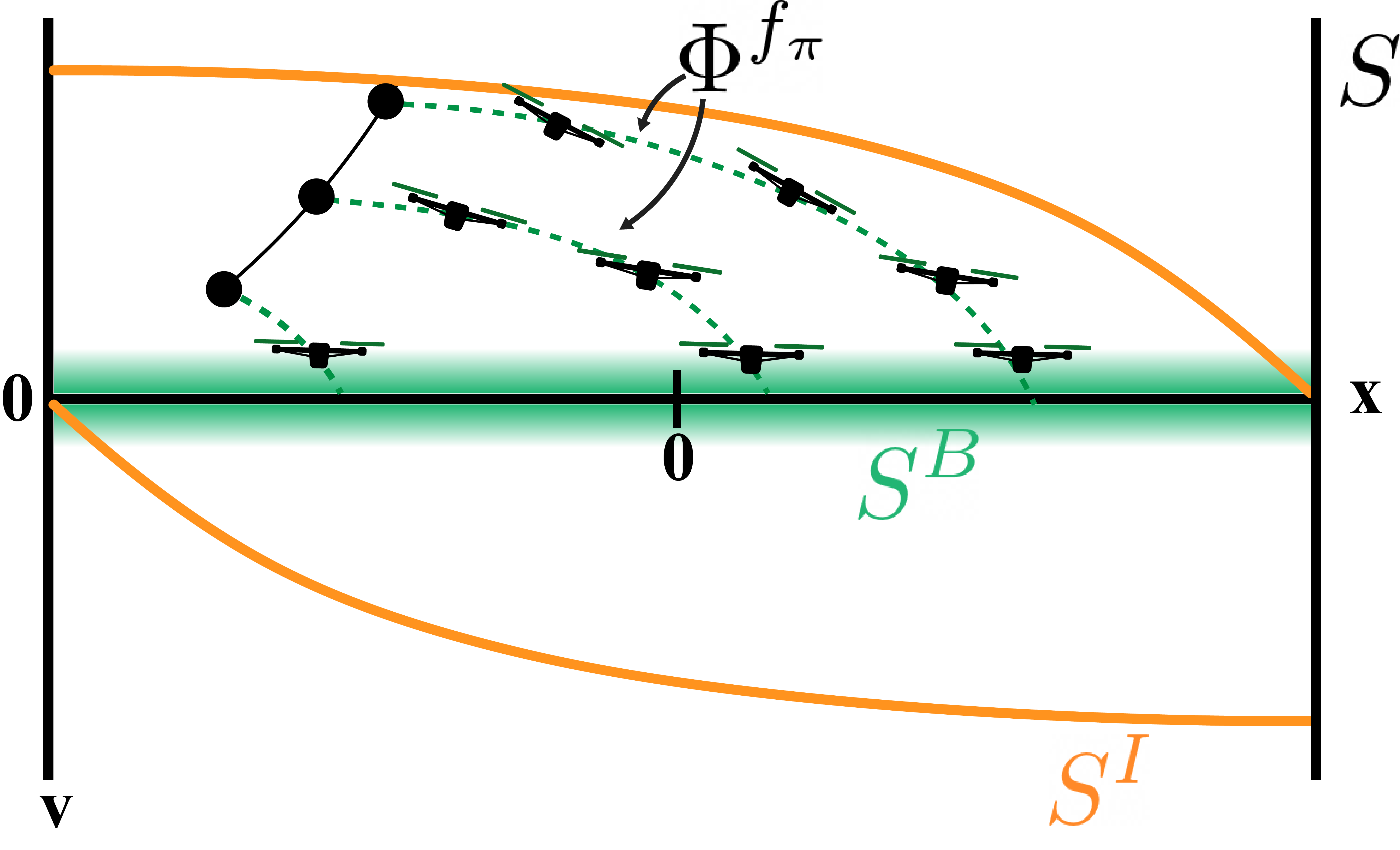}
    \caption{Construction of invariant set $S^I$ by the flow of the backup controller $\Phi^{f_{\pi}}$ to the backup set $S^B$}
    \label{fig:backup_set_illustrated}
\end{figure}

\subsection{Control Filtering with Regulation Functions}

In this setting, we assume there exists a desired control input $\udes$ that we wish to render safe. This means we want a filtering function $\uact(x, \udes) \in U$ that renders a safe set $S$ forward invariant while still remaining as close as possible to $\udes$. This can be achieved using ImCBF's under the assumption that we have a backup controller $u_B$ that renders the set $S_I \subseteq S$ control invariant.
The function $\uact: \R^n \times U \rightarrow U$ must satisfy the following conditions:

\begin{enumerate}[(i)]
    \item The backup controller is applied when $x$ is at the boundary of the safe set: 
    \begin{align}
     h_I(x) = 0 \rightarrow \uact(x,\udes) = \pi(x)   
    \end{align}
    \item $\uact$ is locally Lipschitz continuous in its arguments.
\end{enumerate}

We call any filter $\rho(x,h_I(x),\udes$ that satisfies these conditions a \textit{regulation function}, as it regulates the desired inputs $\udes$ such that safety is guaranteed. An example regulation function that satisfies both of these conditions is
\begin{equation}
\begin{split}
    &\rho(x,h_I(x),\udes) = \\
    &\quad \lambda(x, h_I(x))\udes(x) + \left(1 - \lambda(x, h_I(x)) \right)\pi(x) \label{eq:bcbf_filter}
\end{split}
\end{equation}
such that $\lambda : \R^n \times \R \times U \rightarrow [0,1]$ is given by
\begin{align}
\lambda(x,h_I(x)) = 1-\exp\left(-\beta{\max\{0,h_I(x)\}}\right).
\end{align}
Here, $\beta$ is a tuning parameter that controls how smoothly to mix the backup controller with the desired controller as you approach the boundary. This regulation filter is demonstrated in \cite{singletary2020safety} where it is used to keep a drone inside a geofence at speeds upwards of 100 km/h. While proven to work well in practice for single agents in simple, static environments, this strategy is vulnerable to overly conservative behavior along the boundary of $\S$, and can lead to deadlock in the multi-agent settings. With this work we propose an extension to backup controllers that overcomes deadlocks in the multi-agent setting while strictly increasing the size of $S_I$.

\section{Theory}
\label{sec:theory}



This section introduces the theoretical framework for the usage of backup controllers experimentally in a multi-agent distributed setting.



\subsection{Time-varying backup controller}
\begin{figure}
\centering
  \includegraphics[width=\columnwidth]{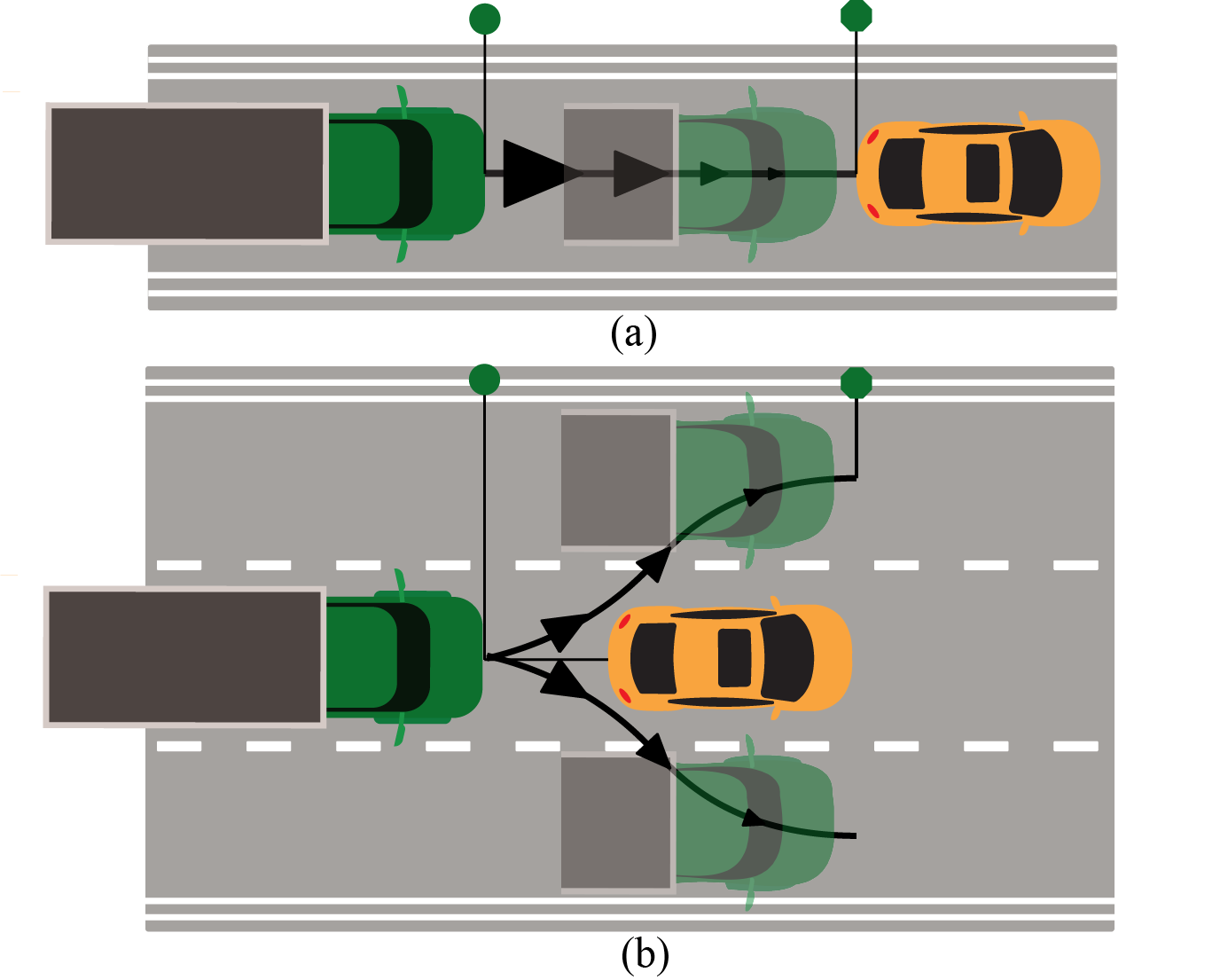}
  \caption{Illustration of the benefits of time-varying backup controllers.  (a) A high-inertia semi truck driving along the highway must keep a large distance behind lighter cars in order to stop before reaching them. (b) By adding a maneuver to switch lanes before stopping, the truck can follow much more closely, under the condition that no one is in the lane. }  \label{fig:car}
\end{figure}

The intuition behind a time-varying backup controller stems from the desire to execute a maneuver before engaging the backup controller that makes it easier for the backup controller to return to the backup set $\S_B$ while staying in $\S$. 
In this section, we will show that even a bad maneuver will not reduce the size of $S_I$ when compared to the original backup controller when properly formulated.
\begin{defn}[Time-Varying Backup Controller]
Let $u_M: \mathbb{R}^n \to U$, $u_B: \mathbb{R}^n \to U$ and $u_{M \rightarrow B} \mathbb{R}^n \times [T_M, T_M+\delta] \to U$
 for some $\delta>0$. We call policies of the following form \textit{time-varying backup controllers}:
\begin{align}
    \pi(x,\tau) = 
   \begin{cases} 
      u_M(x) & \tau \leq T_M \\
      u_{M \rightarrow B}(x,\tau) & T_M \leq \tau \leq T_M + \delta \\ 
      u_B(x) & \tau > T_M + \delta
   \end{cases}. \label{eq:tvbc}
\end{align}
\end{defn}
In \cref{eq:tvbc}, $\pi$ is executing two controllers ($u_M$ and $u_B$) in sequence and continuously transitioning between them with $u_{M \rightarrow B}$. 
The first controller, $u_M$ describes a maneuver to be performed before the backup controller is engaged. 
The controller $u_B$ is the backup controller that brings you to $S_B$.
Finally,  $u_{M \rightarrow B}$ is a time-varying controller that (locally Lipschitz) continuously transitions the input from $u_M(x)$ to $u_B(x)$ over time $\delta$.
The potential benefit from introducing a maneuver is illustrated in Figure \ref{fig:car}.



For a policy $\pi(x,t)$ and initial condition ${x(t_0) = x_{0} \in \mathbb{R}^{n}}$, the solution to this closed-loop system is given by the flow: 
\begin{equation}
    {x(t) = \Phi_{\pi}(x_{0},t)} = x_0 + \int_{t_0}^t \left( f(x) + g(x) \pi(x,\tau) \right)d\tau.
\end{equation}
In this case, for any initial time $t_0 \geq T_M$, the maneuver will not be performed. To remedy this, we introduce an alternative notion for the flow of the system with a time-varying backup controller:
\begin{align}
    x(t)&=\Phi_{\pi}(x_{0},t, \tau_0) \nonumber \\
        &= x_0 + \int_{t_0}^t \left( f(x) + g(x)^\top \pi(x, \tau - \tau_0) \right)d\tau.
        \label{eq:tvbc_flow}
\end{align}
This flow will implicitly maintain forward-invariance of the following set:
\begin{align} \label{eqn:backup_SI_min}
    & S_I(\tau_0) =  \\
   &  \left\{ 
    x \in \R^n \left|  \underset{t \in [0,T]}{\min} \left \{ h\left( \Phi_{\pi}(x,t, \tau_0)\right)
     , h_B\left(\Phi_{\pi}(x,T, \tau_0)\right) 
     \right \} 
     \geq 0\right.
    \right \} \nonumber
\end{align}
which is equivalent to the formulation in \cref{eq:si_min} where the policy $\pi$ corresponds to $\pi$ with the system time offset by $\tau_0$.
The choice of time-offset parameter $\tau_0$ ultimately determines the shape $S_I$. 

\begin{example}
Let us consider two examples that demonstrate how the parameter $\tau_0$ transforms the final invariant set.
First, suppose  $\tau_0 = t_0 - T_M - \delta$. 
In this case,
\begin{align}
    \Phi_{\pi}(x_{0},t, t_0 - T_M - \delta) &= x_0 + \int_{t_0}^t \left( f(x) + g(x)^\top u_B(x) \right)d\tau \nonumber\\
    &= \Phi_{u_B}(x_{0},t) ,
\end{align}
which implies that $S_I(t_0 - T_M - \delta)$ recovers the safe-set of the nominal backup controller $u_B$.

Second, suppose $\tau_0 = t_0 +T - T_M$, which corresponds to the case of only the maneuver being performed over time $T$: $\Phi_{\pi}(x_{0},T, t + T - T_M) = \Phi_{u_M}(x_{0},T)$.  This would likely result in $S_I = \emptyset$, as the backup maneuver $u_M(x)$ would not bring the system back into the backup set $S^B$. 
\end{example}

Clearly, the choice of $\tau_0$ is critical to the performance of the time-varying backup controller and, as demonstrated below, the right choice of $\tau_0$ can yield a control invariant set that is larger than the one guaranteed by the nominal controller. 

\subsection{Time-Offset for Time-Varying Backup Control Filters}

Rather than pick a single $\tau_0$ for the entire evolution of our dynamical system we will allow it to change as a function of the system time and the current state:
\begin{align}
    \tau_0^*(x,t) \coloneqq& \min_{\tau_0 \in [-t, T-t]} \tau_0 \label{eq:optimal_tau} \\
    &\text{s.t.} \quad \Phi_\pi(x,t+\tau, \tau_0) \in S \quad \forall \tau \in [0, T] \label{eq:path_constraint} \\ 
    & \quad \quad ~ \Phi_\pi(x, t+T, \tau_0) \in S_B \label{eq:terminal_constraint}
\end{align}
Recall that in \cref{eq:si_min}, we give the policy a maximum time horizon of $T$ to bring the system into the backup set.
We operate under the assumption that $T > T_M + \delta $ so that the horizon can in principle cover both the maneuver and leave enough time for the backup set controller to take the system back to $S_B$. \Cref{eq:path_constraint} ensures that the chosen $\tau_0^*$ will be safe for states in a horizon of length $T$. The constraint in \cref{eq:terminal_constraint} ensures that the final state of the horizon belongs to the backup set $S_B$. Notice that these two constraints, by definition, imply that $\pi(x,\tau,\tau_0)$ is a valid backup controller with time-offset $\tau_0^*$ since the constraints match the definition in \cref{eq:si_set}. The linear objective ensures we include as much of our maneuver behavior as possible, otherwise we could simply set $\tau_0=t_0 - T_M -\delta$ and only execute the backup controller.

We now show that a time-offset chosen by this optimization problem results in a strict improvement over the nominal case of a single backup controller.

\begin{thm}
\label{thm:strict_improvement}
The safe set for a nominal backup controller $u_B(x)$ is strictly smaller than the safe set provided by a time-varying backup controller $\pi(x,\tau-\tau_0)$ with time-offset $\tau_0^*$ as defined in \cref{eq:optimal_tau} so that the following condition holds true:
\begin{align}
    S_I(u_B) \subseteq S_I(\tau_0^*)
\end{align}
\end{thm}


\begin{proof}
Suppose for the sake of contradiction that there exists an $x \in S_I(u_B)$ so that $x \notin S_I(\tau_0^*)$.  Now $x \in S_I(u_B)$ implies that $\Phi_{u_B}(x,t) \in S \ \forall {t \in [t_0, t_0 +T]}$ and $\Phi_{u_B}(x,t_0+T) \in S_B$ by definition.
Recall that $\tau_0 = t_0 - T_M - \delta$ implies that $$\Phi_{\pi}(x,t, t_0 - T_M - \delta) = \Phi_{u_B}(x,t)$$
Therefore $\tau_0 = t_0 - T_m - \delta$ is feasible for the optimization problem $\tau_0^*$.  This is a contradiction since $x \notin S(\tau_0^*)$ implies $t_0^*$ has no feasible solution but $\tau_0 = t_0 - T_m - \delta$ is a feasible solution.
\end{proof}

Notice that because $\tau_0^*$ is an optimization problem over time rather than state or control input, it is amenable to the time discretization required to numerically integrate ODE's and can be efficiently approximated in practice. We introduce \cref{alg:tau} to approximate $\tau_0^*$ online within the mixing framework described in \cref{eq:bcbf_filter}.

\begin{algorithm}[h!]
\caption{Online Approximation of $\tau_0^*$}\label{alg:tau}
\begin{algorithmic}[1]
\State \textbf{Inputs:}
\State $x$, $t$\Comment{Current State and System Time}
\State $\Delta$ \Comment{Discrete-Time Increment}
\State $\tau_0^*(x(t-\Delta),t-\Delta)$ \Comment{Previous $\tau_0^*$ Solution}
\State $\pi$ \Comment{Time-Varying Backup Controller}
\Procedure{ResetTimeOffset}{}{}
\State $\mathcal{P} \gets \{\Phi_\pi(x,t + \tau,-t) \left| 0 \leq \tau \leq T \right. \}$
\If{$\left(\mathcal{P} \subseteq S \right) \wedge \left( \Phi_\pi(x,t+T,-t) \in S_B\right)$ }
    \State \Return $\tau_0^*(x(t), t) \gets -t$
\Else
    \State \Return $\tau_0^*(x(t), t) \gets \tau_0^*(x(t-\Delta),t-\Delta) + \Delta$
\EndIf
\EndProcedure
\end{algorithmic}
\end{algorithm}


The end result of Algorithm \ref{alg:tau} is that the time-offset $\tau_0$ gets set to the the negation of system time whenever the system is able to perform the backup maneuver $u_M$ for the entirety of $T_M$ and remain safe. Otherwise $\tau_0^*$ is allowed to increase with the system time.
\begin{rem}
The maneuver $u_M(x)$ is chosen not to be time-varying is so that $\tau_0$ may be reset freely before $t-\tau_0 < T_M$ without discontinuities. While there is likely to be a discontinuity in the input when reset after $t-\tau_0 > T_M$, it can only occur every $T_M$ seconds, and the resulting backup maneuver corresponding to the new $\tau_0$ is Lipschitz continuous. Therefore, the reset-induced discontinuities do not affect the safety guarantees, nor do they lead to high-frequency oscillations in the input.
\end{rem}


\subsection{Multiple backup control policies}

Now that the theory is established for the time-varying backup maneuvers, we can introduce the utilization of multiple maneuvers. For a maneuver $i$, the time-varying backup controller takes the following form:
\begin{align} 
    \pi^i(x,\tau) = 
   \begin{cases} 
      u_{M^i}(x) & \tau \leq T_M \\
      u_{M^i \rightarrow B}(x,\tau) & T_M \leq \tau \leq T_M + \delta \\ 
      u_B(x) & \tau > T_M + \delta
   \end{cases}. \label{eqn:multi_maneuver}
\end{align}
For ease of notation, we assume that $T_M$ is constant among the different maneuvers, but in practice, there is no need for this to be the case.

Much like the introduction of a single maneuver $u_M(x)$ resulted in a safe set $S^I$ at least as large as, and generally larger, than the original set, the inclusion of multiple maneuvers can further increase the size of $S^I$. For example, in the case of Figure \ref{fig:car}, having multiple maneuver options could correspond switching to different lanes, increasing the likelihood that one of them is empty.

As shown in \cref{fig:car}, naive mixing of backup maneuvers may not yield a safe control input and could cause discontinuities in the resulting control action taken. In the case of the truck, going left or going right are mutually exclusive. As a secondary consideration we also want to avoid the expensive computation of rolling out trajectories for all possible maneuvers every time we wish to generate a $\uact$.

One possible condition for switching is once it is no longer possible to perform the current maneuver and has engaged only the backup controller portion, i.e. $\tau_0 \geq t_0 - T_M - \delta$. In this case, the backup controller is already engaged, switching to a different policy $\pi^i(x,t,\tau_0)$ will not result in a discontinuity.  This process can be continued for all possible backup maneuvers, until one of the maneuvers allows a reset of $\tau_0$. If no choice of backup maneuver allows a reset, the backup controller continues executing maintaining the safety of the system. This is formalized in \Cref{alg:tau_switch}.

\begin{algorithm}[h!]
\caption{Maneuver Switching Algorithm}\label{alg:tau_switch}
\begin{algorithmic}[1]
\State \textbf{Inputs:}
\State $x$, $t$\Comment{Current State and System Time}
\State $j$ \Comment{Current Maneuver}
\State $\Delta$ \Comment{Discrete-Time Increment}
\State $\tau_0^*(x(t-\Delta),t-\Delta)$ \Comment{Previous $\tau_0^*$ Solution}
\State $\{\pi^i\}_{i=1}^J$ \Comment{Time-Varying Backup Controllers}
\Procedure{SwitchManeuver}{}{}

\If{$\tau_0^*(x(t-\Delta),t-\Delta) \geq t - T_M - \Delta$}
    \State $j \gets j+1$
    \If{$j > J$}
        \State $j = 1$
    \EndIf
        \State $\mathcal{P}^j \gets \{\Phi_{\pi^j}(x,t + \tau,-t) \left| 0 \leq \tau \leq T \right. \}$
        \If{$\left(\mathcal{P}^j \subseteq S \right) \wedge \left( \Phi_{\pi^j}(x,t+T,-t) \in S_B\right)$ }
            \State \Return $\tau_0^*(x(t), t) \gets -t$
        \Else 
            \State \Return $\tau_0^*(x(t), t) \gets \tau_0^*(x(t-\Delta),t-\Delta) + \Delta$
        \EndIf
    
\EndIf
\EndProcedure

\end{algorithmic}
\end{algorithm}

\begin{rem}
Similar to \Cref{alg:tau}, this algorithm results in a discontinuity no more frequent than every $T_M$ seconds. Moreover, only one maneuver is evaluated at each time-step, resulting in minimal computational burden.
\end{rem}

\subsection{Decentralized, multi-agent safety}

For backup CBFs, the decentralized multi-agent case was handled in \cite{chen2020guaranteed}. In this work, we follow a similar approach with our regulation functions along backup maneuvers.

Suppose we have a set of agents $a \in \A$, each with its own independent state $x^{a}\in \R^n$ and nominal safe set defined by continuously differentiable $h^{a}: \R^n \to \R$. We are also concerned with avoiding collisions between agents using pair-wise barrier $h^{ab}: \R^n \times \R^n \to \R$. Combining these two conditions we define a safe set for each agent:
\par\vspace{-3.8mm}
\begin{small}
\begin{align}
\label{eqn:Sa}
    S^{a} = \left\{x_0^{a}\in \R^n \left|   \left( \underset{a}{\bigwedge}  h^{a}(x^{a}) \geq 0 \right) \wedge \left( \underset{a \neq b}{\bigwedge} h^{(a,b)}(x^{a},x^{b}) \geq 0 \right) \right.  \right\}
\end{align}
\end{small}
\noindent This leads to the following result on safety in the decentralized case.  

\begin{thm}[Distributed Safety]
Consider a set of agents $\A$, with each agent $a \in \A$ described by \eqref{eqn:dyn}, i.e., $\dot{x}^{a} = f^{a}(x^{a}) + g^{a}(x^{a}) \pi^{a}(x)$ with each agent following a TBC $\pi^{a}$ using independent time-offsets $\tau_0^{*a}$. Then
\begin{itemize}
    \item For all $a \in \A$ if the initial state $x^{a}_0 \in S^{a}$ then $S^{a} $ in \eqref{eqn:Sa} is safe (forward invariant) so long as each agent has state information, $x^{b}$ for $b \in \A \setminus \{ a \}$.
    \item The global safe set 
    \par\vspace{-3.8mm}
    \begin{footnotesize}
    \begin{align}
        S = \left\{x_0 \in \R^{n \lvert \A \rvert } \left| \left( \underset{a}{\bigwedge}  h^{a}(x^{a}) \geq 0 \right) \wedge \left( \underset{a \neq b}{\bigwedge} h^{ab}(x^{a},x^{b}) \geq 0 \right) \right.  \right\}
    \end{align}    
    \end{footnotesize}
    \noindent is forward invariant where $x_0 \in \R^{n\lvert \A \rvert}$ is the concatenation of all agent states: the global state.
\end{itemize}

\end{thm}










\begin{proof}
Recall that by \cref{thm:strict_improvement}, a TBC using $\tau^{*a}_0$ as offset results in a backup controller that renders $S^{a}$ safe (forward invariant) for each agent $a \in \A$. Clearly, if for all $a \in \A$ $S^a$ is forward invariant then $\cap_{a \in \A}$ is forward invariant. This allows us to perform the following derivation:

\par\vspace{-3.8mm}
\begin{footnotesize}
\begin{align}
    &\cup_{a \in A} S^a= \left\{ x \in \R^{n\lvert \A \rvert} \left|  \underset{a \in A}{\bigwedge} x^{a} \in S^a\right. \right\} \\ 
    &= \left\{ x \in \R^{n\lvert \A \rvert} \left|  \underset{a \in \A}{\bigwedge} h^a(x^a) \geq 0 \wedge \forall_{b \in \A \setminus \{a\}} h^{ab}(x^a, x^b) \geq 0 \right. \right\} \label{eq:apply_sa_defn}\\ 
    &= \left\{x_0 \in \R^{n \lvert \A \rvert } \left| \left( \underset{a}{\bigwedge}  h^{(a)}(x^{(a)}) \geq 0 \right) \wedge \left( \underset{a \neq b}{\bigwedge} h^{(ab)}(x^{(a)},x^{(b)}) \geq 0 \right) \right.  \right\} \label{eq:apply_associativity}\\
    &= S
\end{align}
\end{footnotesize}
\noindent Where \cref{eq:apply_sa_defn} follows from the application of the definition of $S^a$ and \cref{eq:apply_associativity} follows from the application of associativity and commutativity the logic and operator ($\wedge$).
\end{proof}

If multiple agents have multiple maneuvers, the current maneuver must be known by all agents. 
As a consequence of $u_B^i(x)$ being constant in time throughout all maneuvers for an agent $i$, the drones are able to independently update their maneuvers at any time $\tau_0 \geq t - T_M - \delta$.
However, all drones must have knowledge of the new maneuver being utilized, in order to know whether or not they can reset $\tau_0$.

\begin{figure}
\centering
  \includegraphics[width=\columnwidth]{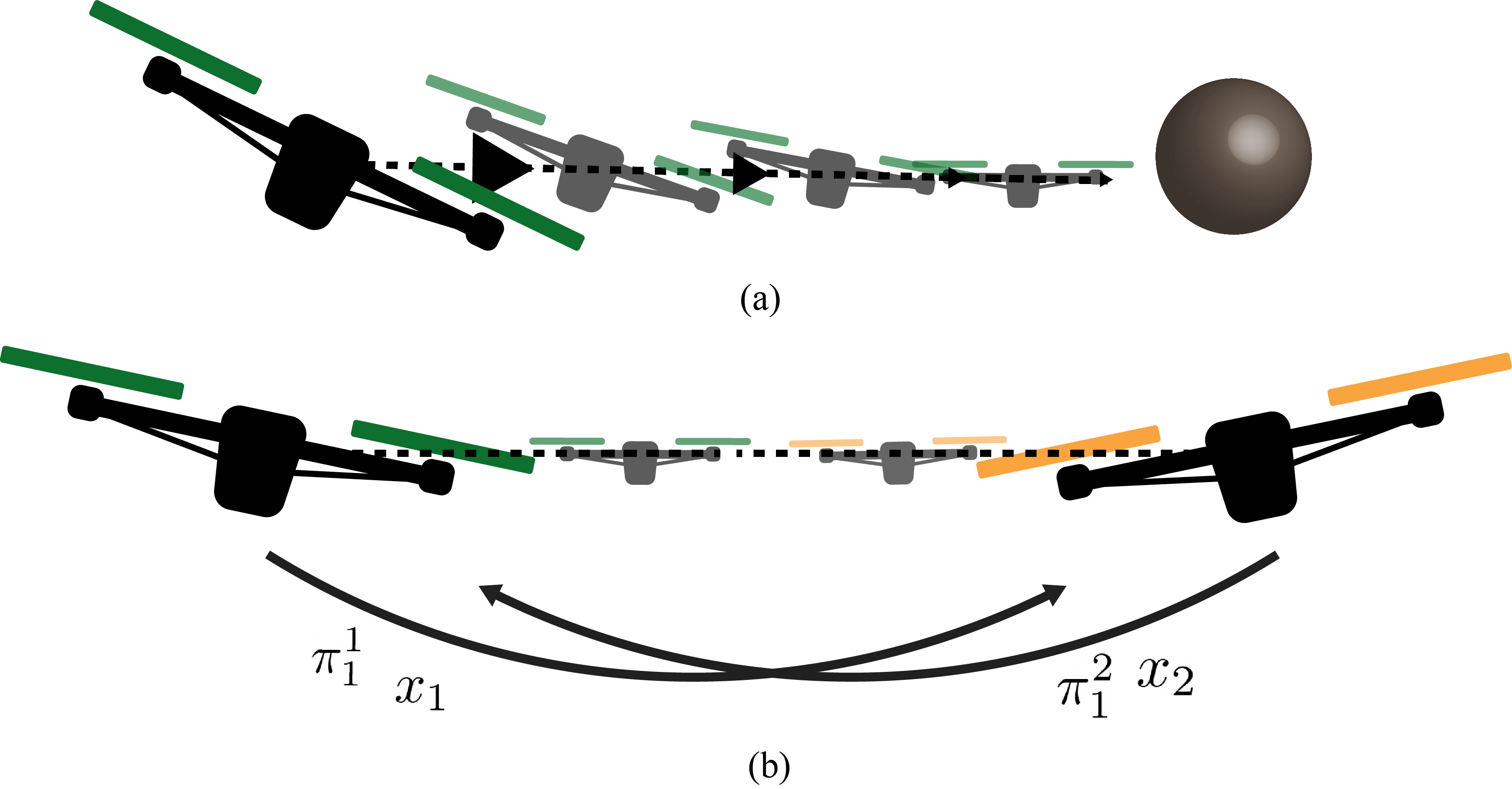}%
  \caption{Illustration contrasting single agent and multi-agent implementation. (a) shows the single agent case, while (b) shows the multi-agent case where state and backup policy information is required to guarantee safety. }  \label{fig:contracts}
\end{figure}

Figure \ref{fig:contracts} demonstrates the idea behind requiring knowledge of the other agents' maneuvers to guarantee safety of the individual.

\section{Modeling and Implementation}
\label{sec:implementation}
\subsection{Drone modeling and control}

By utilizing the high-performance flight controllers on modern racing drones, capable of tracking angular rates at frequencies upwards of 8 kHz, we are able to abstract away the motor and propeller dynamics, and treat the control system as a rigid body in $\R^3$ that attempts to track angular rates. Note that we do not assume perfect angular rate tracking, meaning that this model is still able to capture the dynamics introduced by the motors, propellers, and delays, albeit in a simpler form.

As in the preceding work \cite{singletary2022onboard}, the abstracted system of the drone and lower-level flight controller is modelled as rigid body motion in the Special Euclidean Group in 3 dimensions, SE$(3)$. The state-space model $x \in \R^{13}$ is chosen to be
\begin{equation}
x = [
    p_w ,
    q ,
    v_w ,
    \omega_b]^T,
\end{equation}
where $p_w = [p_x,p_y,p_z]^T$ is the drone position in the world frame, $v_w = [v_x,v_y,v_z]^T$ is the velocity vector in the world frame, $q$ is the the orientation with respect to the world frame represented by a quaternion, and $\omega_b = [\omega_x,\omega_y,\omega_z]^T$ is the angular velocity vector in the body frame.

The angular velocity vector attempts to capture the underlying dynamics via the function
\begin{equation} \label{eqn:omega_model}
    \dot{\omega} = C(x)(\omega_{\textrm{des}}-\omega),
\end{equation}
where $C(x):\R^n \to \R_+$ governs the response time of the tracking controller. 

The throttle-to-thrust curve was fit via a second-order polynomial, using accelerometer and motion capture data. An integral term in the altitude controller is used to eliminate any error in this modeling.

\subsection{Safe sets, backup controller, and regulation function}

For this work, the nominal safe set of each agent $i$ is chosen as a box of free space centered at $(x_c,y_c,z_c) \in \R^3$ with side lengths $r_x,r_y,r_z$,
\begin{equation}
    h^i(x) = \min \left\{r_x^2 - (p_x^i-x_c)^2, r_y^2 - (p_y^i-y_c)^2, r_z^2 - (p_z^i-z_c)^2\right\}.
\end{equation}
The safety constraint between agents $i$ and $j$ is given as
\begin{equation}
    h^{ij}(x) = (p_x^i-p_x^j)^2 + (p_y^i-p_y^j)^2 + (p_z^i-p_z^j)^2 - 4r^2,
\end{equation}
which ensures that no collision occurs between the drones, which are modeled as spheres of radius $r$.

As shown in Section \ref{sec:theory}, the new time-varying backup controllers consist of a maneuver $u_M(x)$ to be performed for time $T_M$, a backup controller $u_B(x)$ to be used after time $T_M + \delta$, and a smoothing controller $u_{M\rightarrow B}(x,t)$ that switches inputs from the two over time $\delta$.

The backup controller $u_B(x)$  is the same velocity controller on SE(3), inspired by \cite{5717652}, that was used in \cite{singletary2022onboard}. The backup controller attempts to stop the drone, and repel it from the boundary of the safe set if it gets too close. 

The backup set $S_B$ is defined by the function
\begin{equation}
    h_B = -\sqrt{v_x^2+v_y^2+v_z^2}+0.1,
\end{equation}
which guarantees that the drone is able to slow itself to a speed of $0.1$ m/s, which is forward invariant under the backup controller.

\subsection{Backup Maneuvers}

\begin{figure*}[t]
    \centering
    \begin{subfigure}{.49\textwidth}
    \includegraphics[trim={0 0 0 0},clip,width=\columnwidth]{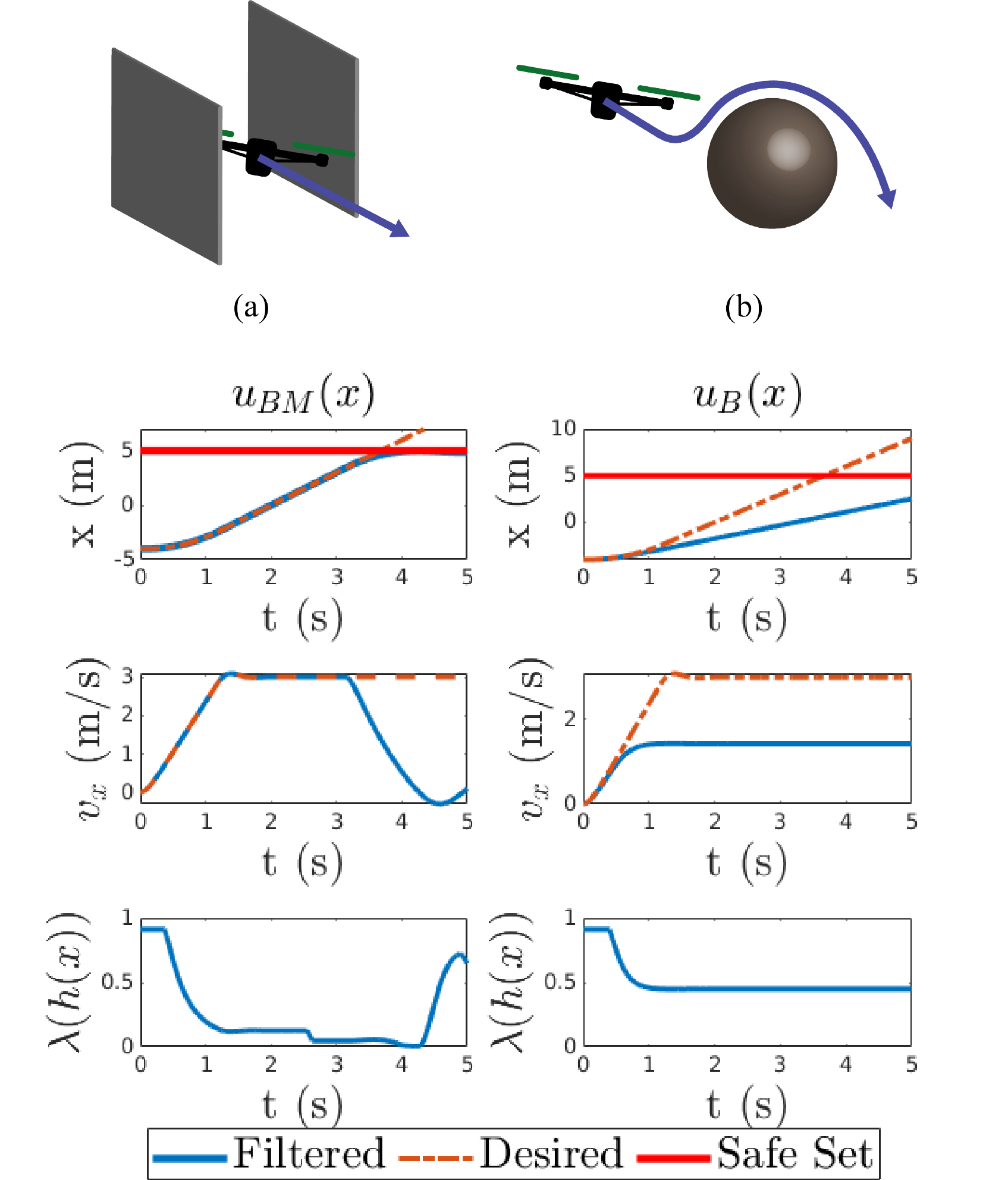}
    \caption{Carry on maneuver compared to only $u_B(x)$ in tightly constrained space.}\label{fig:carryon}
    \end{subfigure}
    \rulesep
    \begin{subfigure}{.49\textwidth}
    \includegraphics[trim={0 0 0 0},clip,width=\columnwidth]{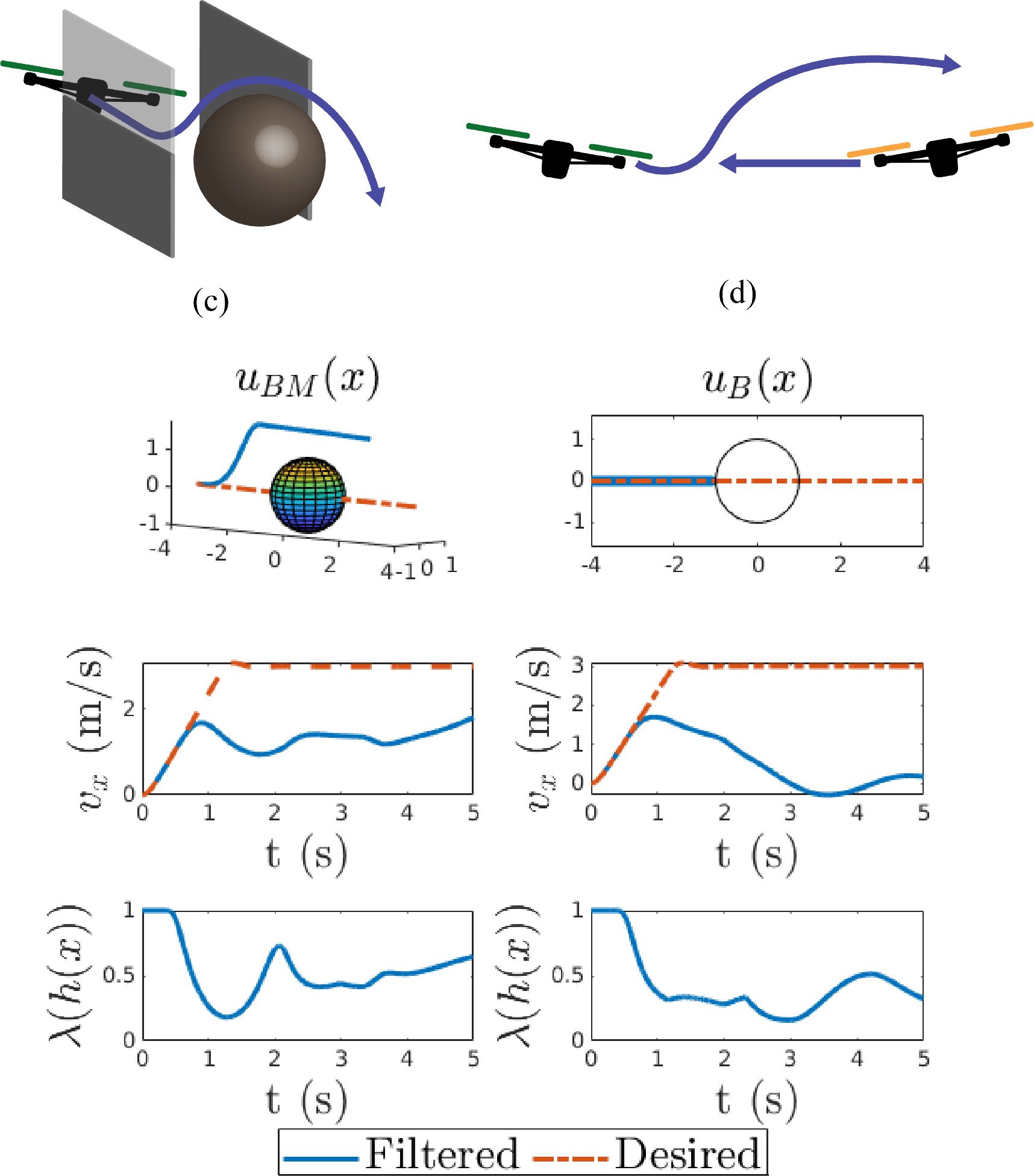}
    \caption{Evade maneuver compared to only $u_B(x)$ when encountering a spherical obstacle. }\label{fig:evade}
    \end{subfigure}
    \begin{subfigure}{.49\textwidth}
    \includegraphics[trim={0 0 0 0},clip,width=\columnwidth]{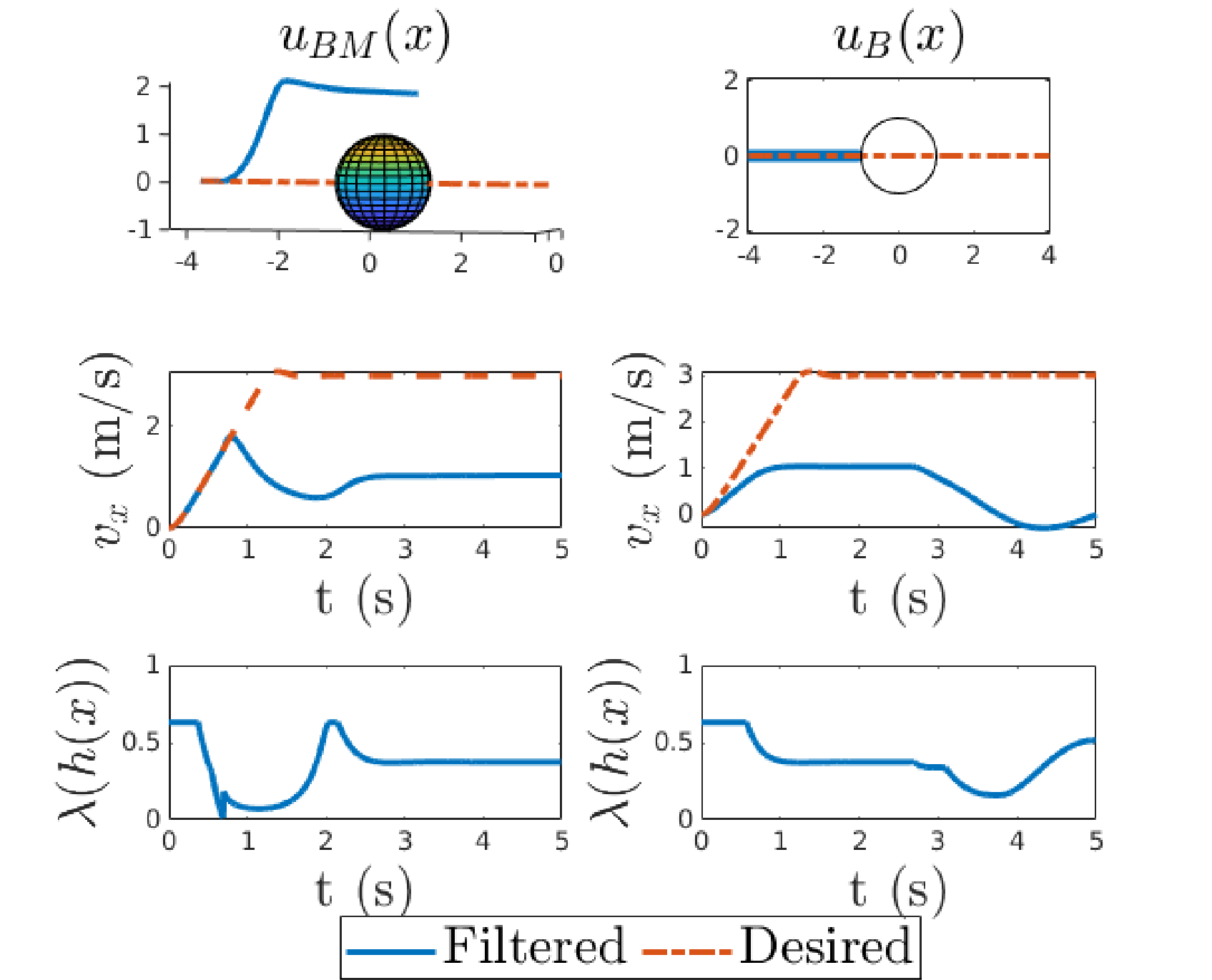}
    \caption{Carry on maneuver switching to evade maneuver when encountering a spherical obstacle in tightly constrained space} \label{fig:switching}
    \end{subfigure}
    \rulesep
    \begin{subfigure}{.49\textwidth}
    \includegraphics[trim={0 0 0 0},clip,width=\columnwidth]{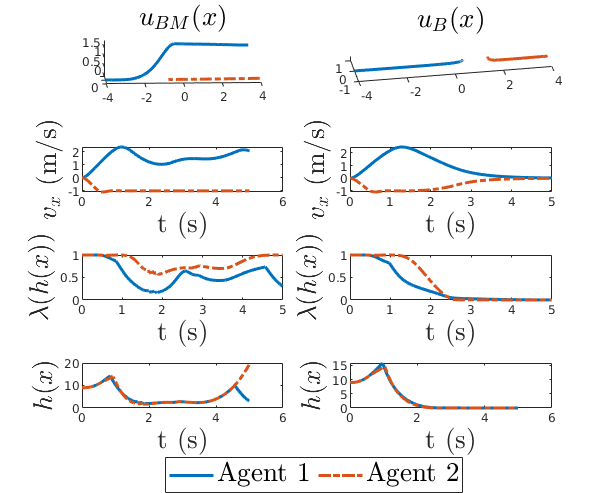}
    \caption{Multi-agent sim} \label{fig:multi-agent-sim}
    \end{subfigure}
    
    \caption{Simulation results capturing the performance benefits of allowing backup maneuvers. While the value of $\lambda$ is often lower, due to being closer or faster near the barrier, there is significantly better alignment with the desired velocities.}
\end{figure*}

In this section, we will outline two simple maneuvers, and demonstrate how they improve performance on hardware. We call these maneuvers the "carry on" and "evade" maneuvers. 

\noindent \textbf{Carry on maneuver.} The carry on maneuver is exactly how it sounds: the maneuver attempts to propagate the current desired input forward. This input is held constant throughout $T_M$, and is updated every time $\tau_0$ is allowed to reset. With the carry on maneuver, the overall policy is defined as
\begin{align} 
    \rho^i(x,\tau) = 
   \begin{cases} 
      \udes & \tau \leq T_M \\ 
      u_{M^i \rightarrow B}(x,\tau) & T_M \leq \tau \leq T_M + \delta \\ 
      u_B(x) & \tau > T_M + \delta
   \end{cases}. 
\end{align}
This policy shines when a skilled pilot wants to move along the edge of the safe set, when one might have $\lambda << 1$. This is illustrated in Figure \ref{fig:carryon}.

\noindent \textbf{Evade maneuver.}
The evade maneuver is a maneuver that attempts to reposition the drone before coming to a stop. This is similar to a lane change, as in \Cref{fig:car}. For this specific maneuver, we attempt to bring the drone upwards, but similar results could be achieved with evading in different directions. The policy is written as
\begin{align} 
    \rho^i(x,\tau) = 
   \begin{cases} 
      u_z(x) & \tau \leq T_M \\ 
      u_{M^i \rightarrow B}(x,\tau) & T_M \leq \tau \leq T_M + \delta \\ 
      u_B(x) & \tau > T_M + \delta
   \end{cases}, 
\end{align}
with $u_z(x)$ being the velocity controller attempting to track to the desired height. This policy shines when the pilot does not react to avoid an obstacle fast enough, and relies on the safety filter to do it for them. This is illustrated in Figure \ref{fig:evade}.

\subsection{Switching backup policies}

By implementing multiple backup policies that the agents can switch between, you can can further increase performance by expanding the size of $S_I$. By combining the carry on and evade maneuvers, and allowing free switching between them in the event that $t \geq T_M$ for the backup maneuver, we achieve the result obtained in Figure \ref{fig:switching}.

\subsection{Multi-agent backup policies}

Finally, the decentralized, multi-agent approach was tested for two agents moving towards each other. Agent 1 is utilizing with the evade maneuver, while Agent 2 attempts the carry-on maneuver. The desired velocity $v_x$ of Agent 1 is 3 m/s, and for Agent 2, it is -1 m/s.

Figure \ref{fig:multi-agent-sim} demonstrates the advantage over simply utilizing the backup controller $u_B(x)$ in preventing the drones from reaching a deadlock.

\section{Hardware Results}
\label{sec:results}
\subsection{Hardware setup}

\begin{figure}
\centering
  \includegraphics[width=\columnwidth]{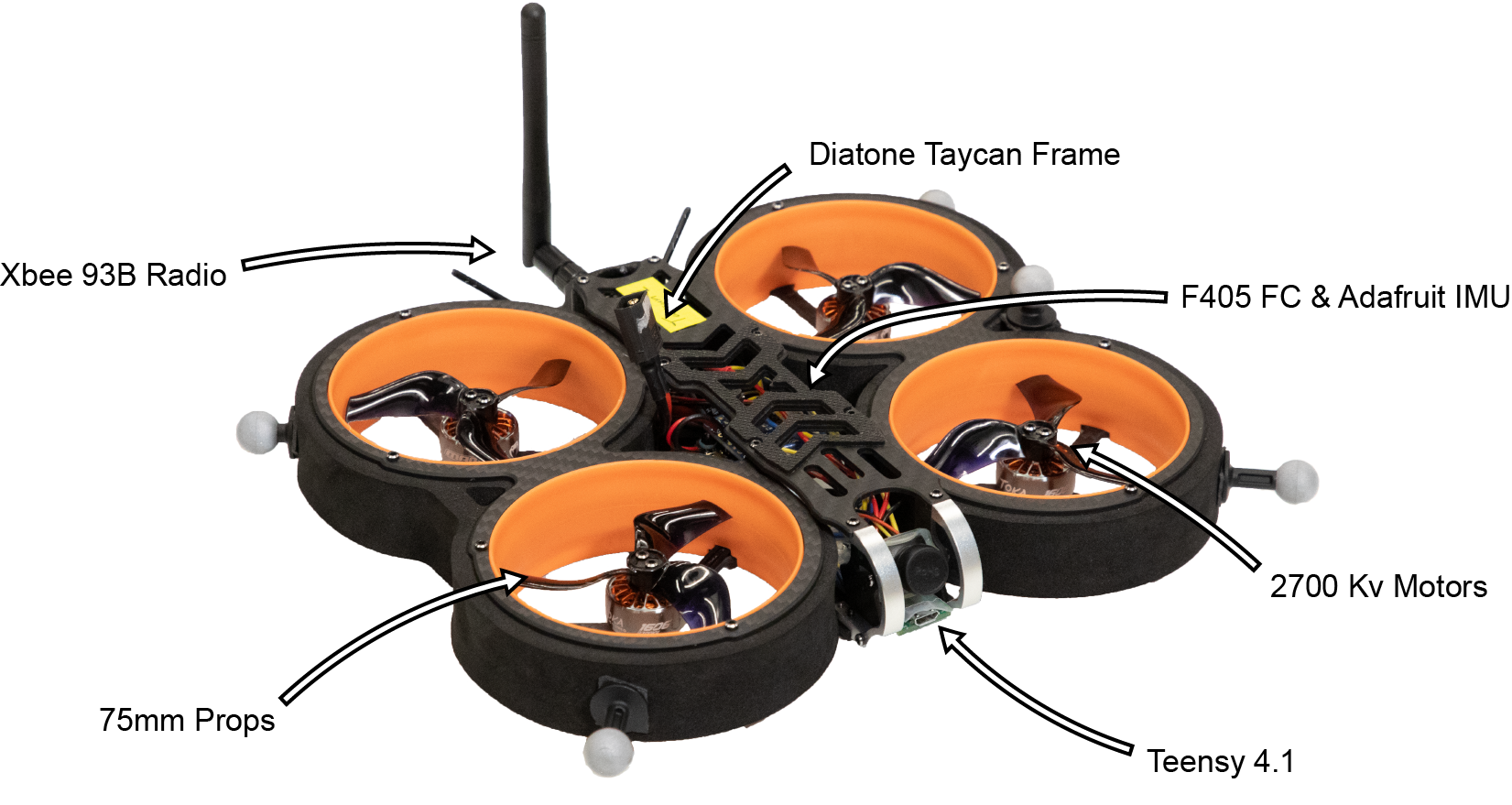}%
  \caption{The Cinewhoop quadrotor used in the experiments}  \label{fig:quad}
\end{figure}
The quadrotor used in our experiments is a consumer "Cinewhoop" drone shown in Fig. \ref{fig:quad}. We add several components to this off the shelf drone for the computation of the barrier functions: a Teensy 4.1 microcontroller, a Xbee 93b radio and an Adafruit BNO055 IMU. Optitrack position and attitude data is streamed to the drone at 20 Hz via a Xbee93B radio. The Teensy 4.1 reads this position data along with IMU data at 100 Hz. The on-board barrier computation issues angular rates commands at 100 Hz to a flight controller running Betaflight, an open source flight control software which is tightly integrated with the flight controller. This allows angular rate tracking at the internal IMU update frequency of 8 kHz. 

\subsection{Hardware results}

\begin{figure}[t!]
\centering
  \includegraphics[width=\columnwidth]{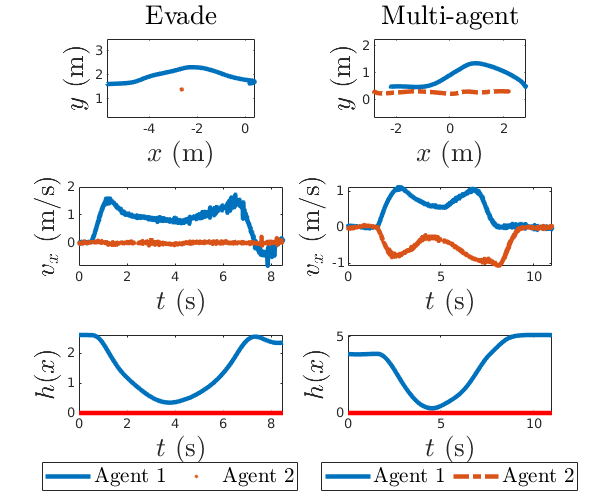}%
  \caption{Plots of the hardware results showcased in Figure \ref{fig:intro}. For more hardware results, please refer to the attached video.}  \label{fig:hardware_results}
\end{figure}

To validate the tractability and robustness of our algorithms in flight, we recreate the simulation results showcased in Figures \ref{fig:evade} and \ref{fig:multi-agent-sim} with two identical "Cinewhoop" drones. The results are demonstrated in Figure \ref{fig:hardware_results}.


The only major difference between the simulation and the hardware was the choice of the evade maneuver to move to the side rather than up. This was done due to limit the effects of the downward airflow on the other agent. As demonstrated in the plots, the TBC's ran quite well despite the very noisy velocity data, as well as the computational constraints of a microcontroller. However, the CBF and all other onboard computations were easily ran at the update rate of the Betaflight's desired inputs, 100 Hz.


\renewcommand{\baselinestretch}{0.98}
\bibliographystyle{IEEEtran}
\bibliography{refs.bib}

\end{document}